\documentclass{revtex4}

\usepackage{graphicx}
\usepackage{amsmath}
\usepackage{amssymb}
\begin{document}

\title{Thick brane solution in the presence of two interacting scalar fields}
\author{Vladimir Dzhunushaliev
\footnote{Senior Associate of the Abdus Salam ICTP}} 
\email{dzhun@krsu.edu.kg} \affiliation{Dept. Phys. and Microel. 
Engineer., Kyrgyz-Russian Slavic University, Bishkek, Kievskaya Str. 
44, 720021, Kyrgyz Republic}


\begin{abstract}
It is shown that two gravitating scalar fields may form a thick  brane in 5D spacetime. 
The necessary condition for the existence of such a regular solution is that the scalar  fields potential must have local and global minima.

Key words:  thick brane, scalar fields
\end{abstract}

\pacs{11.25.-w}
\maketitle

\section{Introduction}

In recent years there has been a revived interest in theories having a greater number of  spatial dimensions than the three that are observed. In contrast to the original  Kaluza-Klein theories of extra dimensions, the recent incarnations of extra dimensional  theories allow the extra dimensions to be large and even infinite in size (in the  original Kaluza-Klein theories the extra dimensions were curled up or compactified 
to the experimentally unobservable small size of the Planck length: $10^{-33}$cm). These new  extra  dimensional theories have opened up new avenues to explaining some of the open  questions in particle physics (the hierarchy problem, nature of the electro-weak symmetry  breaking, explanation of the family structure) and astrophysics (the nature of dark 
matter, the nature of dark energy) \cite{arkani} - \cite{gogberashvili}. In addition they predict new experimentally measurable phenomenon in high precision gravity experiments,  particle accelerators, and in astronomical observations. 
\par
Most of the brane world models use infinitely thin branes with delta-like localization 
of matter. However these models are generally regarded as an approximation since 
any fundamental underlying theory, such as quantum gravity or string theory, must 
contain a fundamental length beyond which a classical space-time description is 
impossible. It is therefore necessary to justify the infinitely thin brane approximation 
as a well-defined limit of a smooth structure -- a thick brane -- obtainable as 
a solution to coupled gravitational and matter field equations. One early example of a  thick brane comes from the 5 dimensional model considered in \cite{rubakov} where one had a  topologically non-trivial field configuration for the scalar field. 
\par 
In Ref. \cite{akama}, the picture is presented that our universe is a dynamically localized 3-brane in a higher dimensional space (''brane world`` ). As an example, the dynamics of the Nielsen-Olesen vortex type in six dimensional spacetime is adopted to localize our space-time within a 3-brane. At low energies, everything is trapped in the 3-brane, and the Einstein gravity is induced through the fluctuations of the 3-brane. 
\par 
It is therefore of great interest to formulate some general requirements on the brane 
world design leading to the appearance of stable, thick branes having a well-defined 
zero thickness limit and able to trap ordinary matter. Taking a physically reasonable  stress-energy tensor it was shown that in 6 dimensions \cite{gogberashvili2} and also higher extra  dimensions \cite{singleton} one can trap all the Standard Model fields using gravity alone. 
\par 
In Ref's \cite{bronnikov} thick brane world models are studied as $\mathbb Z_2$-symmetric domain walls supported by a scalar field with an arbitrary potential $V(\phi)$ in 5D general relativity and it was shown that in the framework of 5D gravity, a globally regular thick brane always has an anti-de Sitter asymptotic and is only  possible if the scalar field potential $V(\phi)$ has an alternating sign. 
\par 
In Ref's \cite{Gremm1} - \cite{barcelo} some properties of brane models was investigated: localization of gravity, graviton ground state, stability and so on. 
\par 
In Ref. \cite{Barbosa-Cendejas:2005kn} a comparative analysis of localization of 4D gravity on a non $Z_2$-symmetric scalar thick brane in both 5-dimensional Riemannian space time and pure geometric Weyl integrable manifold is presented. 
\par 
Multidimensional space-times with large extra dimensions turned out to be very useful when
addressing several problems of  the recent non--supersymmetric string model realization of the Standard Model at low energy with no extra massless matter fields \cite{kokorelis}.
\par
In Ref. \cite{Dzhunushaliev:2003sq} it is shown that two interacting non-gravitating scalar fields with a non-trivial potential may have a regular spherically symmetric solution. This solution shows that one can avoid the Derrick's theorem \cite{derrick} forbidding the existence of regular static solution in the spacetime with the dimension greater 2 for scalar fields if the potential has a local minimum besides global one. This result allows us to assume that the inclusion of gravitation may not destroy the regularity of similar solutions in 5D spacetime. In Ref.~\cite{Bronnikovc} one example of spherically symmetric solution with a gravitating scalar field is given but in contrast with the solution that will be presented here the potential of the scalar field in Ref.~\cite{Bronnikovc} is negative. 
\par 
The goal of this investigation is to show that there exists a new kind of thick brane solutions that is different with thick brane solutions found in Ref's \cite{DeWolfe:1999cp} \cite{Bronnikov:2005bg}. We will show that the asymptotical behavior of one scalar field allow us to offer trapping of Maxwell electrodynamics and spinor fields on the brane. Especially it is necessary to note that the consideration of two scalar fields allow us to obtain the regular thick brane solution with the potential bounded from below. 

\section{Initial equations}

We consider 5D gravity + two interacting fields. The key for the existence of a regular solution here is that the scalar fields potential have to have \emph{local} and \emph{global} minima, and at the infinity the scalar fields tend to a local but \emph{not} to global minimum. \par 
The 5D metric is 
\begin{equation}
	ds^2 = a(y) \eta_{\mu \nu} dx^\mu dx^\nu - dy^2,
\label{sec2-10}
\end{equation}
where $\mu ,\nu = 0,1,2,3$; $y$ is the $5^{th}$ coordinate; 
$\eta_{\mu \nu} = \left\{ +1, -1, -1, -1 \right\}$ is the 4D Minkowski metric. The Lagrangian for scalar fields $\phi$ and $\chi$ is 
\begin{equation}
	\mathcal L = \frac{1}{2} \nabla_A \phi \nabla^A \phi + 
	\frac{1}{2} \nabla_A \chi \nabla^A \chi - V(\phi, \chi)	,
\label{sec2-20}
\end{equation}
where $A= 0,1,2,3,5$. The potential $V(\phi, \chi)$ is 
\begin{equation}
	V(\phi, \chi)	= \frac{\lambda_1}{4} \left(
		\phi^2 - m_1^2
	\right)^2 + 
	\frac{\lambda_2}{4} \left(
		\chi^2 - m_2^2
	\right)^2 + \phi^2 \chi^2 - V_0 ,
\label{sec2-30}
\end{equation}
where $V_0$ is a constant which can be considered as a 5D cosmological constant $\Lambda$. We consider the case when the functions $\phi, \chi$ are 
$\phi(y), \chi(y)$. The 5D Einstein and scalar field equations are 
\begin{eqnarray}
	R^A_B - \frac{1}{2} \delta^A_B R &=& \varkappa T^A_B ,
\label{sec2-40}\\
	\frac{1}{\sqrt{G}} \nabla_A \left( 
		\sqrt{G} G^{AB} \nabla_B \phi
	\right) &=& - \frac{\partial V\left( \phi, \chi \right)}{\partial \phi} ,
\label{sec2-50}\\
	\frac{1}{\sqrt{G}} \nabla_A \left( 
		\sqrt{G} G^{AB} \nabla_B \chi
	\right) &=& - \frac{\partial V\left( \phi, \chi \right)}{\partial \chi} ,
\label{sec2-60}
\end{eqnarray}
where $\varkappa$ is the 5D gravitational constant; $G_{AB}$ is the 5D metric \eqref{sec2-10} and $G$ is the corresponding determinant. After substituting metric \eqref{sec2-10} into Eq's \eqref{sec2-40} - \eqref{sec2-60} we have the following equations 
\begin{eqnarray}
	-3 \frac{a''}{a} - 3 \frac{a'^2}{a^2} &=& \frac{\varkappa}{4} \left[
		\phi'^2 + \chi'^2 + \frac{\lambda_1}{2} \left( 
			\phi^2 - m_1^2
		\right)^2 + \frac{\lambda_2}{2} \left( 
			\chi^2 - m_2^2
		\right)^2 + 2 \phi^2 \chi^2 - 2 V_0 
	\right] ,
\label{sec2-70}\\
	- 6 \frac{a'^2}{a^2} &=& \frac{\varkappa}{4} \left[
		- \phi'^2 - \chi'^2 + \frac{\lambda_1}{2} \left( 
			\phi^2 - m_1^2
		\right)^2 + \frac{\lambda_2}{2} \left( 
			\chi^2 - m_2^2
		\right)^2 + 2 \phi^2 \chi^2 - 2 V_0 
	\right] ,
\label{sec2-80}\\
	\phi'' + 4 \frac{a'}{a} \phi' &=& \phi \left[
		2 \chi^2 + \lambda_1 \left( \phi^2 - m_1^2 \right)
	\right] ,
\label{sec2-90}\\
	\chi'' + 4 \frac{a'}{a} \chi' &=& \chi \left[
		2 \phi^2 + \lambda_2 \left( \chi^2 - m_2^2 \right)
	\right] ,
\label{sec2-100}
\end{eqnarray}
where $\frac{d (\cdots)}{ dy} = (\cdots)'$. Let us introduce the following dimensionless functions $a/\sqrt{\varkappa/6} \rightarrow a$, $\phi \sqrt{\varkappa/3} \rightarrow \phi$, 
$ \chi\sqrt{\varkappa/3} \rightarrow \chi$, 
$2\left( \varkappa/6 \right)^2 V_0 \rightarrow V_0$, 
$m_{1,2}\sqrt{\varkappa/3} \rightarrow m_{1,2}$, 
$\lambda_{1,2}/2 \rightarrow \lambda_{1,2}$ and the dimensionless variable 
$y/\sqrt{\varkappa/6} \rightarrow y$. 
\par 
After algebraical transformations Eq's \eqref{sec2-70} - \eqref{sec2-100} have the following form 
\begin{eqnarray}
	\frac{a''}{a} - \frac{a'^2}{a^2} &=& 
		- \frac{1}{2}\left( \phi'^2 + \chi'^2 \right),
\label{sec2-75}\\
	\frac{a'^2}{a^2} &=& \frac{1}{8} \left[
		\phi'^2 + \chi'^2 - \frac{\lambda_1}{2} \left( 
			\phi^2 - m_1^2
		\right)^2 - \frac{\lambda_2}{2} \left( 
			\chi^2 - m_2^2
		\right)^2 - \phi^2 \chi^2 + 2 V_0 
	\right] ,
\label{sec2-85}\\
	\phi'' + 4 \frac{a'}{a} \phi' &=& \phi \left[
		\chi^2 + \lambda_1 \left( \phi^2 - m_1^2 \right)
	\right] ,
\label{sec2-95}\\
	\chi'' + 4 \frac{a'}{a} \chi' &=& \chi \left[
		\phi^2 + \lambda_2 \left( \chi^2 - m_2^2 \right)
	\right] .
\label{sec2-105}
\end{eqnarray}
It is easy to see that Eq. \eqref{sec2-75} is the consequence of Eq. \eqref{sec2-85}: if we take a derivative from the LHS and RHS of Eq. \eqref{sec2-85} then we shall receive Eq. \eqref{sec2-75}. The boundary conditions are 
\begin{eqnarray}
	a(0) &=& a_0 ,
\label{sec2-110}\\
	a'(0) &=& 0 ,
\label{sec2-120}\\
	\phi(0) &=& \phi_0 , \quad \phi'(0) = 0 , 
\label{sec2-130}\\
	\chi(0) &=& \chi_0 , \quad \chi'(0) = 0 . 
\label{sec2-140}
\end{eqnarray}
The boundary condition \eqref{sec2-120}-\eqref{sec2-140} and Eq. \eqref{sec2-85} give us the following constraint 
\begin{equation}
	V_0 = \frac{\lambda_1}{4} \left( 
			\phi^2_0 - m_1^2
		\right)^2 + \frac{\lambda_2}{4} \left( 
			\chi^2_0 - m_2^2
		\right)^2 + \frac{1}{2} \phi^2_0 \chi^2_0,
\label{sec2-150}
\end{equation}

\section{Numerical investigation}
\label{num}

For the numerical calculations we choose the following parameters values 
\begin{equation}
	a_0 = \phi_0 = 1, \quad 
	\chi_0 = \sqrt{0.6}, \quad 
	\lambda_1 = 0.1, \quad 
	\lambda_2 = 1.0 .
\label{sec3-10}
\end{equation}
We apply the methods of step by step approximation for finding of numerical solutions using the MATHEMATICA package (the details of similar calculations can be found in Ref. \cite{Dzhunushaliev:2003sq}, the corresponding MATHEMATICA program can be found in *.tar.gz file of the archived version of this paper \cite{Dzhunushaliev:2006vv}). 
\par 
\textbf{Step 1}. On the first step we solve Eq. \eqref{sec2-105} (having zero approximations $a_0(y) = a_0, \chi_0(y) = m_1 \tanh y$). The regular solution exists for a special value $m^*_{1,i}$ only. For $m_1 < m^*_{1,i}$ the function $\chi_i(y) \rightarrow +\infty$ and for $m_1 > m^*_{1,i}$ the function $\chi_i(y) \rightarrow -\infty$
(here the index $i$ is the approximation number). One can say that in this case we solve \emph{a non-linear eiqenvalue problem}: $\chi_i^*(y)$ is the eigenstate and $m_{1,i}^*$ is the eigenvalue on this Step. 
\par 
\textbf{Step 2}. On the second step we solve Eq. \eqref{sec2-95} using zero approximation $a_0(y)$ for the function $a(y)$ and the first approximation $\chi_1^*(y)$ for the function $\chi(y)$ from the Step 1. For $m_2 < m^*_{2,1}$ the function $\phi_1(y) \rightarrow +\infty$ and for $m_2 > m^*_{2,1}$ the function $\phi_1(y) \rightarrow -\infty$. Again we have \emph{a non-linear eiqenvalue problem} for the function $\phi_1(y)$ and $m^*_{2,1}$.
\par 
\textbf{Step 3}. On the third step we repeat the first two steps that to have the good convergent sequence $\phi_i^*(y), \chi_i^*(y)$. Practically we have made three approximations. 
\par 
\textbf{Step 4}. On the next step we solve Eq. \eqref{sec2-75} which gives us the function $a_1(y)$. 
\par 
\textbf{Step 5}. On this step we repeat Steps 1-4 necessary number of times that to have the necessary accuracy of definition of the functions $a^*(y), \phi^*(y), \chi^*(y)$. 
\par
\begin{figure}[h]
\begin{minipage}[t]{.45\linewidth}
  \begin{center}
  \fbox{
  \includegraphics[height=5cm,width=7cm]{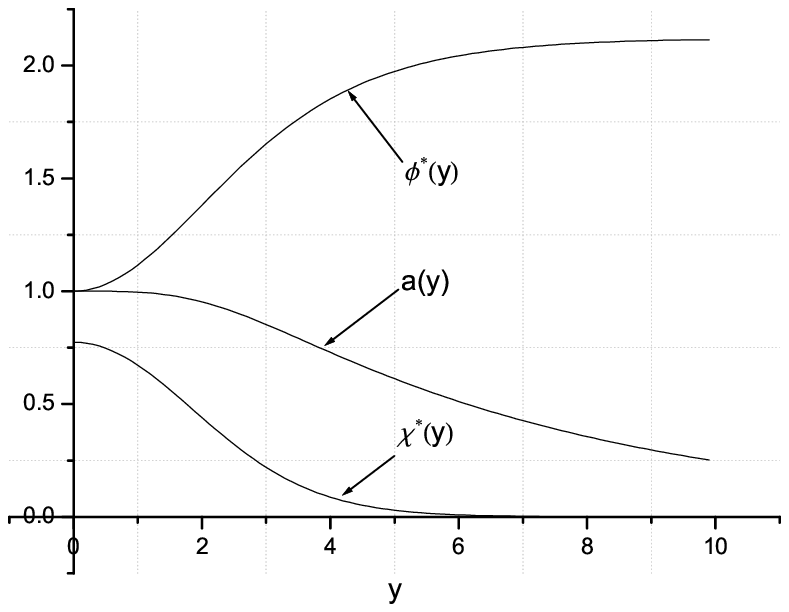}}
  \caption{The functions $a^*(y), \phi^*(y), \chi^*(y)$}
  \label{fig1}
  \end{center}
\end{minipage}\hfill
\begin{minipage}[t]{.45\linewidth}
  \begin{center}
  \fbox{
  \includegraphics[height=5cm,width=7cm]{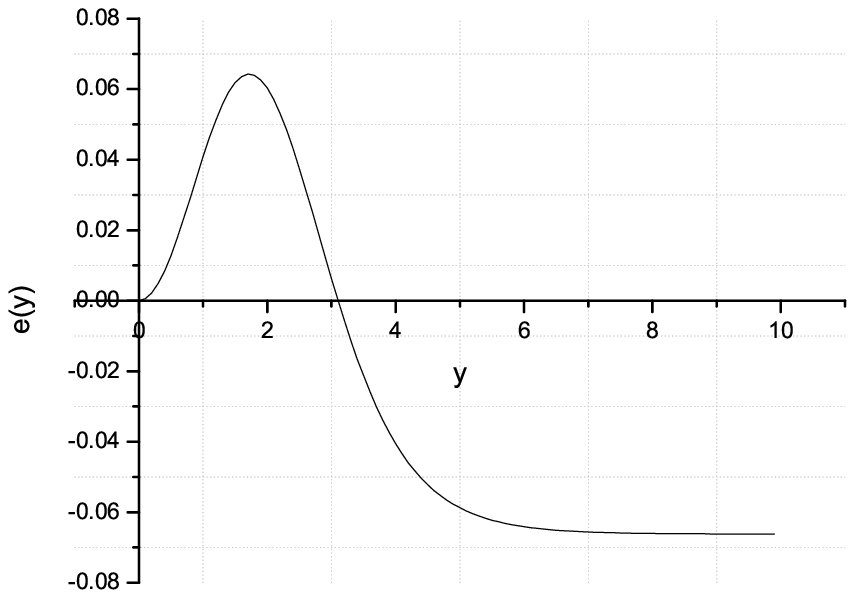}}
  \caption{The profile of dimensionless energy density.}
  \label{fig2}
  \end{center}
\end{minipage}\hfill 
\end{figure}

 
After Step 5 we have the solution presented on Fig. \ref{fig1}. These numerical calculations give us the eigenvalues $m_1^* \approx 2.122645756$, 
$m_2^* \approx 1.3721439906788$ and eigenstates $a^*(y), \phi^*(y), \chi^*(y)$. The derived solution was verified by using the standard numerical method of solving the differential equations in the MATHEMATICA package (the corresponding MATHEMATICA program can be found in *.tar.gz file of the archived version of this paper \cite{Dzhunushaliev:2006vv}). 
\par 
It easy to see that the asymptotical behavior of the solution is 
\begin{eqnarray}
	a(y) &\approx& a_\infty e^{-k_a y} , \quad
	k_a^2 = \frac{1}{4} \left( V_0 - \frac{\lambda_2}{4} m_2^4 \right) ,
\label{sec3-20}\\
	\phi(y) &\approx& m_1 + \phi_\infty e^{-k_\phi y}  ,\quad 
	k_\phi = 2k_a + \sqrt{4 k_a^2 + 2 \lambda_1 m_1^2}  ,
\label{sec3-30}\\
	\chi(y) &\approx& \chi_\infty e^{-k_\chi y}  ,\quad 
	k_\chi = 2k_a + \sqrt{4 k_a^2 + m_1^2 - \lambda_2 m_2^2}  ,
\label{sec3-40}
\end{eqnarray}
where $a_\infty, \phi_\infty, \chi_\infty$ are constants. The dimensionless energy density is 
\begin{equation}
	e(y) = 2 \left( \frac{\varkappa}{3} \right)^2 \varepsilon(y) = 
	\frac{1}{4} \left[
		\phi'^2 + \chi'^2 + \frac{\lambda_1}{2} \left( \phi^2 - m_1^2 \right)^2 + 
		\frac{\lambda_2}{2} \left( \chi^2 - m_2^2 \right)^2 + 
		\phi^2 \chi^2 - 2 V_0
	\right]
\label{sec3-50}
\end{equation}
and it is presented in Fig. \ref{fig2}. 
\par 
Taking into account that the quantity 
$V\left( \phi(\infty), \chi(\infty) \right)$ is absolutely similar to a 5D cosmological constant, we can introduce a dimensionless brane tension 
\begin{equation}
	\sigma = 2 \int \limits_0^\infty \biggl[ 
		e(y) - V\Bigl( \phi(\infty), \chi(\infty) \Bigl)
	\biggl] dy \approx 0.74 .
\label{sec3-55}
\end{equation}
\par
\begin{figure}[h]
  \begin{center}
  \includegraphics[height=9cm,width=9cm]{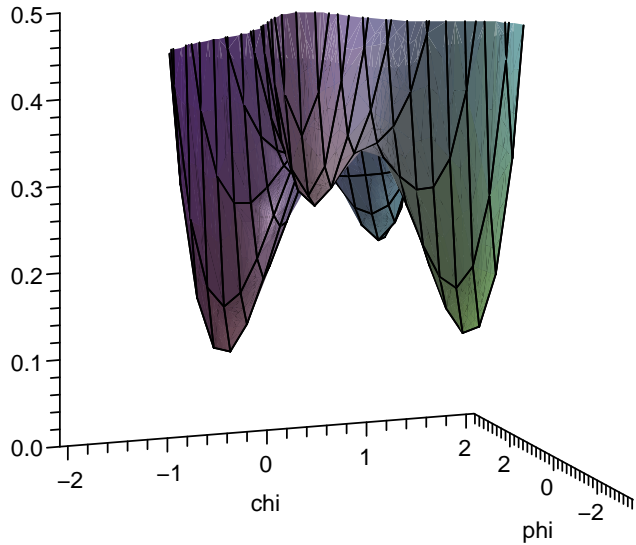}
  \caption{The profile of the potential $V(\phi, \chi)$.}
  \label{fig3}
  \end{center}
\end{figure}

 
According to Eq. \eqref{sec3-30} \eqref{sec3-40} one can define the thickness $\Delta$ of the presented  thick brane as 
\begin{equation}
	\Delta \approx \max \left\{ k_\phi, k_\chi \right\}.
\label{sec3-60}
\end{equation}
The key role for understanding why such regular solution may exist belongs to the fact that the potential \eqref{sec2-30} has the local and global minima. The profile of the potential $V(\phi, \chi)$ is presented in Fig. \ref{fig3}. 

\section{Trapping of the matter}

Now we would like to consider trapping of the electromagnetic and spinor fields on the above  derived thick brane. The Lagrangian of interacting electromagnetic and scalar fields is taken from :
\begin{equation}
\label{Lagr_int}
	L_{eff}=-\frac{1}{4} \tilde{F}_{BC} \tilde{F}^{BC} + 
	\alpha \phi^2 \tilde{A}_B \tilde{A}^B - m^2 \tilde{A}_B \tilde{A}^B ,
\end{equation}
where $\tilde{F}_{BC} = \tilde{A}_{B,C} - \tilde{A}_{C,B}$ is the 5D electromagnetic tensor with 5-dimensional vector potential $\tilde{A}_B(x^{D})$ and scalar field $\varphi(y)$ depending only on the extra coordinate $y$; $\alpha$ - an arbitrary constant, $m$ is the mass of vector field $\tilde{A}_B$.
\par
\begin{figure}[h]
\begin{center}
\fbox{
  \includegraphics[height=6cm,width=8.4cm]{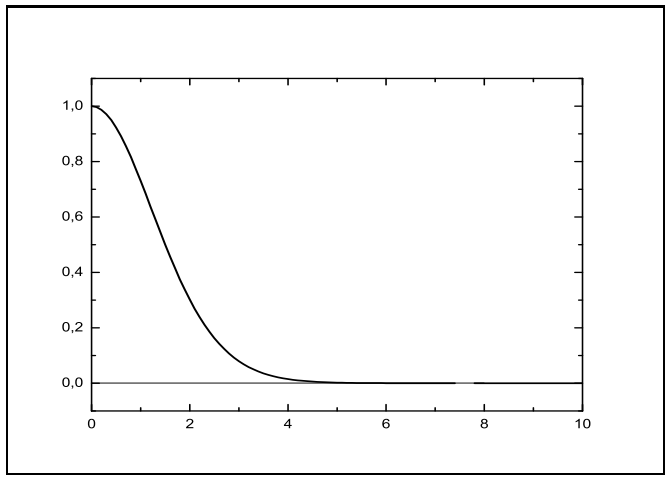}}
 \caption{The function $f(r)$ with $\alpha=2$ and $m \approx 1.6256513943551$}
\label{EM}
\end{center}
\end{figure}

The 5D Maxwell equations will be:
\begin{equation}
\label{max}
    D_C \tilde{F}^{BC}=\tilde{A}^B(x^{D}) \left[ \alpha \phi^2 - m^2 \right].
\end{equation}
Let us rewrite Eq. (\ref{max}) as follows:
\begin{equation}
\label{max1}
    D_{\nu} \tilde{F}^{B \nu}+D_5 \tilde{F}^{B 5}=
    \tilde{A}^B(x^{D}) \left[ \alpha \phi^2 - m^2 \right].
\end{equation}
We will use the gauge $\tilde{A}_5=0$ and search for a solution of (\ref{max1}) in the form:
\begin{eqnarray}
    D_\nu \tilde{F}^{\mu \nu} &=& 0,
\label{max2}\\
    D_5 \tilde{F}^{B 5} &=& \tilde{A}^B(x^{D}) \left[ \alpha \phi^2 - m^2 \right]
\label{max3}
\end{eqnarray}
For the solution we will use the following ansatz
\begin{equation}
    \tilde{A}^B(x^{D}) = A^B(x^{\mu}) f(y),
\end{equation}
where $A^B(x^{\mu})$ is the 4D electromagnatic potential function only on 4D coordinates. Then from Eq's \eqref{max2} \eqref{max3} we will have
\begin{eqnarray}
  D_\nu F^{\mu \nu} &=& 0,
\label{sec4-10}\\
  f^{\prime \prime} + 4\frac{a^{\prime}}{a} f^{\prime} &=& 
  \frac{1}{a^4} \frac{d}{dy} \left( a^4 \frac{df}{dy} \right) = 
  f \left( \alpha \phi^2 - m^2 \right)
\label{sec4-20}
\end{eqnarray}
where the first equation is the usual 4D Maxwell equations on the brane. The solution of the second equation on the background of the thick brane is presented in Fig.\ref{EM}. Here it is necessary to note that again the regular solution $f(y)$ exists for an exceptional value of the parameter $m$ only. It is easy to see from Eq.~\eqref{sec4-20}: this equation is exactly  Schrodinger equation with the potential $\alpha \phi^2$ (which is a hole). Eq.~\eqref{sec4-20} has a regular solution describing a particle in a hole for an exceptional value of $m$ that is an eigenvalue of the Schrodinger equation \eqref{sec4-20}. 
\par 
As one can see, the EM field is trapped on the 4D brane. In this case the  electromagnetic fields in the bulk are
\begin{equation}
    \tilde{A}(x^B) = A(x^\mu) f(y)
\end{equation}
where $f(y)$ is the exponentially decreasing function.

Let us consider further the question about trapping of fermion fields on the brane. In the
simplest case such a possibility was pointed out in Ref.~\cite{Rub} at consideration of the brane model as the model of domain wall. In this work the model of one real scalar field $\phi$ with two degenerated minima was introduced for description of the domain wall in 5D spacetime. In this case existing kink solution has its asymptotes in these minima with constant values of the field $\phi$. In our case similar situation occurs: two scalar fields $\phi, \chi$ create the system with two local minima, and the solutions tend asymptotically to one of these minima where the field $\chi$ tends to zero and $\phi$ to the constant values as in the case from Ref.~\cite{Rub}.
\par 
It allows us to investigate trapping of fermions on the brane for our case by analogy with Ref.~\cite{Rub}. The curved space 5D gamma matrices are
\begin{equation}
    \Gamma^{\mu} = \frac{1}{\sqrt{a}} \gamma^{\mu},\qquad \Gamma^{r}=-i \gamma^{5},
\end{equation}
where $\gamma^{\mu}$ and $\gamma^{5}$ are the usual Dirac matrices in 4D theory
\begin{equation}
	\gamma^\mu = \left\{ \left(
	\begin{array}{cc}
		0 & I \\
		I & 0
		\end{array} 
	\right), 
	\left(
	\begin{array}{cc}
		0 						& \vec \sigma \\
		-\vec \sigma 	& 0
		\end{array} 
	\right)
	\right\}, \quad 
	\gamma^5 =  \left(
	\begin{array}{cc}
		-1 & 0 \\
		0  & 1
		\end{array} 
	\right)
\label{1.3} 
\end{equation} 
where $\sigma^i$ are usual Pauli matrices in flat spacetime. Then using the action for interacting scalar $\varphi$ and fermion $\Psi$ fields we have 
\begin{equation}
\label{action_int}
    S_{\Psi}=\int d^4x\, dr \left( i \overline{\Psi} \Gamma^A D_A \Psi -
    h \phi \overline{\Psi} \Psi \right)
\end{equation}
here $\phi$ is the scalar field from the Lagrangian \eqref{sec2-20} and $h$ is a constant. The Dirac equation can be written in the form
\begin{equation}
\label{Dirac_eq}
 i \Gamma^A D_A \Psi-h \phi(y) \Psi =0.
\end{equation}
Here $D_A=\partial_A+\Upsilon_A$, where pseudo-connection $\Upsilon_A$ can be defined as follows~\cite{Ying}
$$
\Upsilon_A=\frac{1}{2} e_{\bar{M}}^N \left(\partial_A e_N^{\bar{M}}-\partial_N e_A^{\bar{M}}\right),
$$
where the vielbein $e_A^{\bar{M}}$ is defined via $g_{A B}= e_A^{\bar{M}} e_B^{\bar{N}} \eta_{\bar{M} \bar{N}}$, and the inverse vielbein 
$e_{\bar{M}}^A$ via $g^{A B}= e_{\bar{M}}^A e_{\bar{N}}^B \eta^{\bar{M} \bar{N}}$.
For our case $\Upsilon_A=\left(0, 0, 0, 0, a^{\prime}/a\right)$.
\par 
Let us consider ansatz
\begin{equation}
\label{ansatz1}
    \Psi (x^B) = \psi(x^\mu) \Psi_0(y)
\end{equation}
If we are interested in localization of zero modes, then, as it was shown in Ref.~\cite{Jack},
there are the solutions of Eq. (\ref{Dirac_eq}) with 4D mass $m=0$. For the zero mode $\gamma^{\mu}D_{\mu}\psi=0$,
and the Dirac equation (\ref{Dirac_eq})  turns out in the equation:
\begin{equation}
\label{Dirac_eq_1}
    \gamma^5 \left( \partial_r +\frac{a^{\prime}}{a}\right) \Psi_0 = 
    h \phi(y) \Psi_0,
\end{equation}
where $^\prime$ means the derivative with respect to $r$. Eq. \eqref{Dirac_eq_1} with account of \eqref{ansatz1}
has the following solution:
\begin{equation}
\label{Dirac_sol}
    \Psi = \exp{\left[ -\int_{0}^{r} dr^{\prime}\left(\frac{a^{\prime}}{a}+ h 											\phi(r^{\prime})\right) \right]}
    \psi (x^\mu),
\end{equation}
where $\psi (x^\mu)$ is the usual solution of 4D Weyl equation, and the condition 
$\gamma_5 \Psi_0=-\Psi_0$ is taken into account. As it was shown in Section \ref{num}, the sum $\left(a^{\prime}/a+ h \phi\right)$ tends asymptotically to some constant. So the zero mode (\ref{Dirac_sol}) is localized near $r=0$, i.e. on the brane, and decreases exponentially at large $r$: $\Psi_0 \propto \exp{(-m_5 \left| y \right|)}$. 
\par 
Let us note that one can include the function $\chi$ in Eq's \eqref{Lagr_int} and \eqref{action_int} by the following way: \\ 
$\alpha \phi^2 \tilde{A}_B \tilde{A}^B \stackrel{\text{change}}{\longrightarrow} 
\alpha \left( \phi^2 + \chi^2 \right) \tilde{A}_B \tilde{A}^B$ and 
$h \phi \overline{\Psi} \Psi \stackrel{\text{change}}{\longrightarrow} 
h \left( \phi + \chi \right) \overline{\Psi} \Psi$ but it does not matter because the asymptotical behavior of the function 
$\phi \stackrel{r \rightarrow \infty}{\longrightarrow} m_1$ is important only 
(as $\chi \stackrel{r \rightarrow \infty}{\longrightarrow} 0$). 
\par 
Let us note that we do not consider trapping of scalar fields on this brane. The reason is very simple: we have shown \emph{exactly} that two scalar fields with Lagrangian \eqref{sec2-20} are confined on the brane. The situation is even better: these scalar fields create the brane ! It is necessary note that in the process of numerical calculation we have obtained a domain wall solution without gravity, i.e. two scalar fields can create the solution with the planar symmetry and switching on the gravity does not destroy this solution. 

\section{Discussion and conclusions}

Now we would like to list the essential specialities of the presented solution:
\begin{enumerate}
	\item The existence of the solution crucially depends on the number of interacting scalar fields $(n > 1)$ and the presence of the non-trivial potential $V(\phi, \chi)$ which has local and global minima. At the infinity the scalar fields tend to local minimum and the potential has alternating sign when $r \in [0, \infty]$ that to the existence of the presented solution. The numerical investigation shows that in the presence of one scalar field the similar solution does not exist. 
	\item The advantage of the presented solution is that the asymptotical behavior \eqref{sec3-30} of the scalar field $\phi$ allow us to obtain trapping of electromagnetic and spinor fields on the brane. 
	\item Let us note that the thick brane solution presented here differs from the thick brane solutions presented in Ref's \cite{DeWolfe:1999cp} and \cite{Bronnikov:2005bg} that: 
	\begin{enumerate}
		\item thick brane solution from Ref.~\cite{DeWolfe:1999cp} is obtained for the scalar field with the potential unbounded from below that in contrast with our potential \eqref{sec2-30} which is bounded from below.
		\item in Ref.~\cite{Bronnikov:2005bg} the thick brane solution is obtained for scalar fields having non-trivial asymptotical topological structure in contrast with our solution. 
	\end{enumerate}	
	\item The solution is topologically trivial. It means that at the infinity two scalar fields do not form a hedgehog configuration in contrast with the thick brane solutions presented in Ref.~\cite{Bronnikov:2005bg}.
	\item The quantity $V(\phi(\infty), \chi(\infty))$ can be considered as a 5D cosmological constant $\Lambda$. 
	\item \label{4} In Ref. \cite{Dzhunushaliev:2003sq} it is shown that after some simplification and assumtions the SU(3) gauge Lagrangian can be reduced to the Lagrangian \eqref{sec2-20} describing interacting scalar fields $\phi$ and $\chi$. This remark allows us to assume that a real thick brane can be formed by a 5D gauge condensate which is described by interacting scalar fields. 
	\item According to the previous item (\ref{4}) the 5D mechanism of trapping the matter on a thick brane may be similar to the confinement mechanism in 4D quantum chromodynamics. In this case trapping of the corresponding quantum gauge fields on the thick brane is non-perturbative and can not be investigated using Feynman diagram technique. 
	\item If the thick brane is formed with the help of a gauge condensate then the problem of the stability of the thick brane becomes very non-trivial. It occurs because a non-static condensate has to be described in much more complicated manner than static condensate in Ref. \cite{Dzhunushaliev:2003sq}. It is connected to that fact that the change in time of quantum object is connected not only to the change of this quantity but also to the change of a wave function as well. 
	\item From the mathematical point of view the presented solution is an eigenstate for a nonlinear eigenvalue problem. 
\end{enumerate}

\end{document}